\documentclass[10pt,prb,aps,twocolumn,showpacs,superscriptaddress,floatfix]{revtex4}

\usepackage{graphicx}
\usepackage{bm}
\usepackage{amssymb}
\usepackage{color,ulem}

\begin{document}
\title{Origin of the Phonon Hall Effect in Rare-Earth Garnets}

\author{Michiyasu Mori}
\affiliation{Advanced Science Research Center, Japan Atomic Energy Agency,Tokai 319-1195, Japan}

\author{Alexander Spencer-Smith}
\affiliation{School of Physics, University of Sydney, Sydney 2006, Australia}

\author{Oleg P. Sushkov}
\affiliation{School of Physics, University of New South Wales, Sydney 2052, Australia}

\author{Sadamichi Maekawa}
\affiliation{Advanced Science Research Center, Japan Atomic Energy Agency,Tokai 319-1195, Japan}

\begin{abstract}
The phonon Hall effect has been observed in the paramagnetic insulator, Tb$_3$Gd$_5$O$_{12}$. 
A magnetic field applied perpendicularly to a heat current induces a temperature gradient that is
perpendicular to both the field and the current. 
We show that this effect is due to resonant skew scattering of phonons from the crystal field states
of superstoichiometric Tb$^{3+}$ ions. 
This scattering originates from the coupling between the quadrupole moment of Tb$^{3+}$ ions and the lattice strain. 
The estimated magnitude of the effect is consistent with experimental observations at 
$T\sim5$ K, and can be 
significantly enhanced by increasing temperature.
\end{abstract}

\date{\today}

%66.70.-f	Nonelectronic thermal conduction and heat-pulse propagation in solids; thermal waves (for electronic thermal conduction in metals and alloys, see 72.15.Cz and 72.15.Eb)
%72.20.Pa	Thermoelectric and thermomagnetic effects
%72.15.Gd	Galvanomagnetic and other magnetotransport effects (see also 75.47.-m Magnetotransport phenomena; materials for magnetotransport)

\maketitle
%\section{Introduction}
When a linear magnetic field is applied perpendicularly to a heat current in a sample of terbium gallium garnet (TGG), Tb$_3$Ga$_5$O$_{12}$, 
a transverse temperature gradient is induced in the third perpendicular direction~\cite{strohm,inyushkin07}. 
This is the \lq\lq phonon Hall effect (PHE)". 
The effect was observed in this insulator at low temperature ($T\sim5$ K), a situation in which there are no mobile charges such as electrons or holes~\cite{note0}. 
The Neel temperature of TGG is $0.24$ K~\cite{hammann}, so it is a paramagnet at $T\sim5$ K.
Hence magnons do not contribute to the heat current and one does not expect a  contribution from 
the magnon Hall effect~\cite{onose08,katsura,onose10,Mats11}.
Phonons are not charged and hence cannot be affected by the Lorentz force which gives rise to the usual
classical Hall effect. 
Therefore the mechanism for the PHE must be related to the spin-orbit interaction.
However, the spin-orbit interaction for phonons is not at all obvious, unlike in the 
anomalous Hall effect and spin Hall effect for electrons~\cite{Dyak71,Hirs99,Maek06}. Thus, an understanding of the origin of the observed PHE is a fundamental problem.

So far, there have been a few theoretical attempts to explain the PHE~\cite{sheng,kagan,maksimov,Shi12}.
Refs.~\onlinecite{sheng} and \onlinecite{kagan} assumed a Raman-type interaction between the spin of
stoichiometric Tb$^{3+}$ ions and the phonon. This interaction results in \lq\lq elliptically polarized'' phonons. According to \onlinecite{sheng,kagan}, the \lq\lq elliptic polarization'', in combination with scattering from impurities, leads to the PHE. In this scenario the type of impurity is unimportant and hence phonon -- impurity scattering is considered in the leading Born approximation.
This is an intrinsic-extrinsic scenario, 
i.e.,  
the \lq\lq elliptic polarization'' is 
an intrinsic effect and the scattering from impurities is an extrinsic effect.
The major problem with this scenario was realized in Ref.~\onlinecite{maksimov} - in spite of the 
\lq\lq elliptic polarization'' the Born approximation does not result in the PHE. 
Ref.~\onlinecite{maksimov} attempted to go beyond the leading Born approximation in impurity scattering. 
However, the problem has not been resolved yet. 
An intrinsic mechanism for the PHE, based on the Berry curvature of phonon bands, was suggested in
Ref.~\onlinecite{Shi12}. This is similar to the Berry curvature mechanism in the Hall effect for 
light~\cite{Onod04}. The Berry curvature mechanism is certainly valid
for materials with specially structured phonon bands, however, it is hard to see how the 
mechanism can be 
realized in TGG which has the simple cubic structure.

There is an important experimental observation which was missed in all the previous theoretical
analyses of the PHE - TGG crystals can be grown by the flux method (TGG$_{fl}$),  and by the Czochralski method 
(TGG$_{G}$). While TGG$_{fl}$ has perfect stoichiometry, TGG$_{G}$ contains about 1\% of 
superstoichiometric Tb$^{3+}$ ions. 
At 5 K the diagonal thermal conductivity of
TGG$_{G}$ is about 5 times smaller than that of TGG$_{fl}$~\cite{inyushkin10}.  
This indicates that the thermal conductivity in TGG$_{G}$ is determined
by phonon scattering from the crystal field states of
superstoichiometric Tb$^{3+}$ ions~\cite{inyushkin10}.
The PHE has only been observed in TGG$_{G}$~\cite{strohm,inyushkin07}.
Thus, one concludes that the PHE is of extrinsic origin - due to the phonon scattering
from superstoichiometric Tb$^{3+}$ ions.
We stress that the PHE in TGG relies specifically upon
scattering from superstoichiometric Tb$^{3+}$ ions, not just scattering from any impurities.
This observation was not considered in all previous suggestions~\cite{sheng,kagan,Shi12} for the mechanism behind the PHE.

In this Letter, motivated by the above observation, we show that the PHE originates from the resonant 
skew scattering of phonons from the crystal field states of superstoichiometric Tb$^{3+}$ ions. 
Below, we will often refer to superstoichiometric Tb$^{3+}$ ions as impurities.

{\it Phonons}.--
The phonon Lagrangian density reads
\begin{eqnarray}
\label{L0}
{\cal L}_0&=&
	\frac{\rho}{2}\left\{{\dot{\varphi}}_j^2-c_T^2(\partial_i{\varphi}_j)^2
-(c_L^2-c_T^2)(\partial_i{\varphi}_i)(\partial_j{\varphi}_j)\right\}\nonumber,\\
{\bm{ \varphi}}&=&
	\sum_{{\bm {q}},\mu}\frac{{\bm e}_{\mu}}{\sqrt{2\rho\omega_{{\bm{ q}}\mu}}}
\left[a_{{\bm{ q}}\mu}e^{-i\omega_{{\bm{ q}}\mu}t+i{\bm{ q}}\cdot{\bm{ r}}}+ h.c.\right]. \label{quant}
\end{eqnarray}
Here ${{\bm{ \varphi}}}$ is lattice displacement.
The isotropic model (\ref{L0}) is known to be appropriate for a system with a large unit cell, such as that of a 
garnet~\cite{kittel,Plan77,note3}. 
 The index $\mu=1,2,3$ enumerates phonon polarization, ${\bm{ e}}^{(\mu)}$ is the
unit polarization vector, $a_{{\bm{ q}}\mu}$ is the annihilation operator of the phonon, and 
$\omega_{{\bm{ q}}\mu}=c_Lq~(c_Tq)$ is the energy of the longitudinal (transverse) phonon.
For the purpose of making estimates, we will use the following value of speed: 
$c \approx 
3.7 \times 10^5$ cm/s, and the mass density: $\rho=7.2$ g/cm$^3$~\cite{inyushkin10}.
Below, only the longitudinal mode is considered. 
It is plausible that this mode dominates PHE due to its large velocity, $c_L \approx 2 c_T$ \cite{Plan77}. Even if transverse modes gave comparable contribution,
this does not influence our estimate of the effect.

{\it Tb ion}.--
The $^7$F$_6$ state of a free 
Tb$^{3+}$ ion splits into 
13 levels in the dodecahedral crystal field of the garnet. 
The energies of low lying levels in intrinsic ions
are approximately 0, 3, 49, 62, 72, 76 K \cite{Koningstein, Hammann}.
The energy levels of impurity ions (superstoichiometric) depend on their particular positions,
but overall they are comparable to those of ions in regular sites.
The thermal conductivity in TGG$_G$ is mainly determined by the resonant scattering of phonons
from superstoichiometric ions.
Note that resonant scattering necessarily implies a nonzero scattering phase 
and hence gives rise to skew scattering, which does not appear in the Born approximation~\cite{LL}.

Fitting the measured diagonal thermal conductivity~\cite{inyushkin10} within four levels of the impurity ion, 
we come to the ion level scheme shown in Fig.~\ref{fig1} left,
$\omega_{ab}=3$K, $\omega_{ac}=20$K, $\omega_{ad}=70$K~(see supplemental material).
It is known that the ground state energy doublet is very sensitive to
magnetic field $B$.
At $T=4.2$K and $B < 1$T the ion magnetic moment grows linearly
with $B$. At fields larger than 1-2 T, the magnetic moment 
practically saturates at $|M| \approx 4~\mu_B$~\cite{kolmakova90a}.
This data indicates that the a,b-states are composed of time conjugate states $|\pm M\rangle$,  
$|a\rangle\propto |+M\rangle+|-M\rangle$,
$|b\rangle\propto |+M\rangle-|-M\rangle$,
and, subjected to a under magnetic field, the a,b-states evolve to $|\pm M\rangle$ as shown in
Fig.~\ref{fig1} right, $\omega_{ab} \to \Omega_{a'b'}=\sqrt{\omega_{ab}^2+(2gB)^2}$
with an effective $g$ factor~\cite{note4}.
%%%%%%%%%%%%%%%%%
\begin{figure}[ht]
\includegraphics[width=0.35\textwidth]{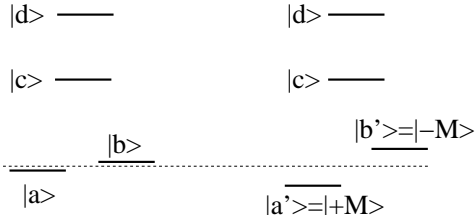}
\caption{The crystal field levels for $B=0$ on the left,
and for $B > 1-2$ T on the right. 
}
\label{fig1}
\end{figure}
%%%%%%%%%%%%%%%%%
Below, we assume that, for the magnetic field larger than 1-2 T, only the state $|a'\rangle=|+M\rangle$ is thermally populated,
while, without magnetic field, both $|a\rangle$ and $|b\rangle$ are populated. 
For simplicity we assume that $|c\rangle$ and $|d\rangle$ are not sensitive to the magnetic field.

{\it Spin-phonon interaction}.--
The quadrupole Coulomb interaction of a Tb ion with its surrounding lattice ions is of the following
form~\cite{abragam,Fulde},
\begin{eqnarray}
\label{L1}
H_1&=&
\gamma T_{ij}\partial_i\varphi_j \\
T_{ij}&=&\frac{3}{2J(2J-1)}\left\{J_iJ_j+J_jJ_i-\frac{2}{3}J(J+1)\delta_{ij}\right\} \ . \nonumber
\end{eqnarray}
Here $\varphi_j$ is the lattice displacement at the ion site $i,j=x,y,z$. 
The quadrupole moment 
$Q_{ij}=QT_{ij}$ is written in terms of the total angular
momentum $J$. This implies that the strong spin orbit interaction inside
the ion core is embedded in Eq.~(\ref{L1}).
The size of an ion core is about one Bohr radius $a_B$. Hence,
the quadrupole moment $Q$ is roughly estimated as $Q\sim ea_B^2$, where $e$ is
the elementary charge.
The gradient of the  electric field $E$ from the surrounding ions is estimated as 
$\nabla E\sim e/d^3$, where 
$d \approx 2 \AA$
is the distance  to the nearest oxygen ion. Then, the magnitude of the coupling $\gamma$ is,  
\begin{equation}
\label{gg}
\gamma\sim Q\nabla E \sim  \frac{e^2a_B^2}{d^3} \sim  0.7~{\rm eV} \ .
\end{equation}

{\it Resonant scattering}.--
Phonon scattering from  superstoichiometric Tb$^{3+}$ ions
is determined by the diagram in Fig.~\ref{fig2}. 
%%%%%%%%%%%%%%%%
\begin{figure}[ht]
\includegraphics[width=0.35\textwidth]{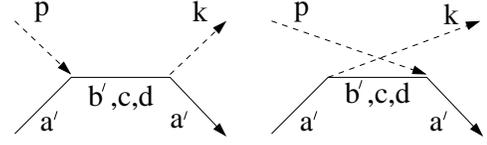}
\caption{Amplitude of phonon scattering from a Tb ion with virtual 
excitation of the crystal field level given magnetic field larger than 1-2 T. 
The solid line shows the ionic state and 
the dashed-line shows the phonon. 
Without the magnetic field, the initial state $a'$ is substituted by the states $a$ or $b$, with $c$ and $d$ as intermediate states. 
}
\label{fig2}
\end{figure}
%%%%%%%%%%%%%%%%%
Under nonzero magnetic field, a straightforward calculation gives the following scattering rate
for a phonon with energy $\omega$,
\begin{eqnarray}
\label{w1m}
&&\tau^{-1}_{\omega}= \tau_L^{-1}+\sum_{i=b',c,d}\tau^{-1}_{a'i,\omega},\nonumber\\
&&\tau^{-1}_{a'i,\omega}=\frac{N_s}{N_{Tb}} \frac{\omega_D^3\omega^4}{80\pi}
 \frac{(\Omega_{a'i}/\omega_{ai})^2\Gamma_{ai}^2/\omega_{ai}^4}
{(\omega^2-\Omega_{a'i}^2)^2+\Omega_{a'i}^2\Gamma_{i \omega}^2},\nonumber\\
&&\Gamma_{ai}=\gamma^2\omega_{ai}^3/\pi\rho c^5\ , \ \ \ \ 
\Gamma_{i \omega}=(\omega/\omega_{ai})^3\Gamma_{ai} \,
\end{eqnarray}
Here $\tau_{L}^{-1}= c/L$ is due to the finite size of the  sample $L\approx 1$mm. 
The total density of Tb ions is $N_{Tb}\approx 1.3\times 10^{22} {\rm cm}^{-3}$, 
the density of superstoichiometric Tb ions is $N_s\approx 1.5\times 10^{20}{\rm cm}^{-3}$, 
and the Debye frequency/temperature is 
$\omega_D =487$ K~\cite{inyushkin10}.
Eq.~(\ref{w1m}) is similar
to that derived a long time ago in Refs.~\onlinecite{Sheard,Toombs}.
It is worth noting that the $\omega^4$ dependence in the numerator of the resonant part of
$\tau_{\omega}^{-1}$ originates from the derivative in the interaction (\ref{L1}).
This derivative is enforced by Adler's theorem~\cite{adler}.

{\it Skew component}.-- We take the magnetic field directed along the z-axis. 
The phonon propagates in the xy-plane with an initial momentum ${\bm k}=k(1,0,0)$   
and final momentum ${\bm q}=k(\cos\phi,\sin\phi,0)$,  where $\phi$ is the scattering angle.
When the magnetic field is small, the states $a'$ and $b'$ are populated and then the diagrams in Fig.~\ref{fig2} give the following phonon angular distributions for scattering~(see Ref.~\onlinecite{note1} and supplemental material),
\begin{eqnarray}
W_{\bm k\to \bm q}^{a'c}
&\approx&\frac{\tau_{a'c,\omega}^{-1}}{2\pi} 
\left( \cos^2 \phi  - 
\frac{\omega \Gamma_{c \omega}}{\Omega_{a'c}^2}
\cos\phi \sin \phi  \right)\ ,\label{w1}\\
W_{\bm k\to \bm q}^{b'c}
&\approx&\frac{\tau_{b'c,\omega}^{-1}}{2\pi} 
\left( \cos^2 \phi  + 
\frac{\omega \Gamma_{c \omega}}{\Omega_{b'c}^2}
\cos\phi \sin \phi  \right). \label{w11} 
\end{eqnarray}
Note that the second term proportional to $\sin\phi$ is the skew component and  
the sign of the $a'c$ process is opposite to that of the $b'c$ process.
This is due to the time-conjugation of the states $|a'\rangle=|+M\rangle$ and $|b'\rangle=|-M\rangle$. 
Without the magnetic field, these process cancel each other out, 
whereas with a magnetic field the skew component becomes finite for two reasons - the energy difference between $\Omega_{a'c}$ and $\Omega_{b'c}$, and the de-population of the state $b'$. 
The $a'b'$ and $b'a'$ processes also contribute to scattering such as 
Eqs.~(\ref{w1}) and (\ref{w11}), respectively.
If the states $a'$ and $b'$ are equally populated, the skew components in these processes cancel each other out, since $\tau_{a'b',\omega}=\tau_{b'a',\omega}$ and $|\Omega_{b'a'}|=|\Omega_{a'b'}|$. 
When the state $b'$ is depopulated by increasing the magnetic field, the cancellation becomes imperfect and the $a'b'$ process also contributes to the skew scattering.
 
{\it Correlation of impurity positions}.-- The $ \cos \phi \sin \phi$ term in 
Eqs.~(\ref{w1}) and (\ref{w11}) change sign
at $\phi \to -\phi$. This is the skew asymmetry which is necessary for the PHE.
However, this term also changes sign  at $\phi \to \pi-\phi$.
Because of this, the off-diagonal thermal conductivity is zero, 
$\kappa_{xy}=0$, in spite of the skew
since skew scattering in the forward hemisphere, $\cos \phi >0$,
is exactly compensated for by skew scattering in the backward hemisphere, $\cos \phi < 0$.
There is no such problem for electron skew scattering~\cite{Fert},
but there is a similar problem for the skew scattering of light.
There are two mechanisms which destroy the $\phi \to \pi-\phi$ 
compensation: 
(i) Spatial correlation of impurity positions discussed below; 
(ii) Interference between contributions with different
values of $\Delta J_z$, this mechanism is discussed in the supplemental 
material.

A superstoichiometric Tb$^{3+}$ ion has ionic radius 0.92{\AA} 
and it replaces a Ga$^{3+}$ ion with smaller radius 0.62{\AA}.
Hence the crystal lattice around the Tb ion is elastically deformed 
towards larger lattice spacing. During the process of crystal growth this creates more room for another superstoichiometric Tb ion in the 
vicinity of the first one. Hence the impurity density $\rho_s({\bm r})$ must be correlated as
\begin{equation}
\label{dd}
\overline{\rho_s(0)\rho_s({\bm r})}=N_s\delta({\bm r})+N_s^2[1
+C e^{-r/l}],
\end{equation}
where the correlation length is about the average distance between impurities, 
$l\approx N_s^{-1/3} \approx 2\times 10^{-7}$ cm.
Given the significant difference in ionic radii it is natural to assume
about a 50\% change in the probability of having  another 
superstoichiometric Tb ion in the  vicinity of the first one. Hence, it is 
reasonable to expect that the correlation constant is 
$C \sim \pm 1$.
Due to the correlation (\ref{dd}), the interference between
phonon scattering amplitudes from adjacent impurities is nonzero
and the scattering probability Eq.~(\ref{w1}) is modified by an interference term as:  $W_{k \to q}\to W_{k \to q}(1+C P_\phi)$, where 
$P_\phi= 1/[1+(2kl\sin\phi/2)^2]^2$.
Thus, the correlation destroys the $\phi \to \pi-\phi$ compensation factor. 
It is convenient to expand $P_\phi$ in  series
of Legendre polynomials $P_\phi=a_0(\omega)+a_1(\omega)P_1(cos\phi)+...$, where
\begin{eqnarray}
a_1(\omega)&=& \frac{3}{(\omega/\omega_1)^2}
\left[
1+\frac{1}{1+(\omega/\omega_1)^2}\right]\nonumber\\
&-&\frac{6}{(\omega/\omega_1)^4}\ln\left[1+(\omega/\omega_1)^2\right],
\end{eqnarray}
and 
$\omega_1\equiv \hbar c/2l \approx 13$K.
Hence, accounting for the mechanisms (i) (see also supplemental material),
%$P_\phi$, 
the scattering rate given by Eqs.~(\ref{w1}) and (\ref{w11}) is transformed to
\begin{eqnarray}
\label{w2}
&&W_{\bm k\to \bm q} \approx \frac{\tau_{\omega}^{-1}}{4\pi} \left\{1  - 
{\cal K}_{\omega}
\omega\Gamma_{c\omega} 
{\bm n_B}\cdot[{\bm n_k}\times{\bm n_q}]\right\}, \\
&&{\cal K}_{\omega}
=
\frac{C}{5}a_1(\omega)\tau_{\omega}
\left(\frac{\tau_{a'c,\omega}^{-1}}{{{\Omega^2_{a'c}}}}
-n_T\frac{\tau_{b'c,\omega}^{-1}}{{{\Omega^2_{b'c}}}}
+\bar{n}_T\frac{\tau_{a'b',\omega}^{-1}}{{{\Omega^2_{a'b'}}}}
\right), \nonumber\\
&&n_T\equiv \exp[-\Omega_{a'b'}/T]\equiv 1-\bar{n}_T,\nonumber
\end{eqnarray}
where ${\bm n_B}, {\bm n_k}, {\bm n_q}$ are unit vectors along the direction of the magnetic field and the phonon momenta respectively, and $n_T$ and $\bar{n}_T$ are the thermal populations.

\begin{figure}[t]
\includegraphics[width=0.45\textwidth]{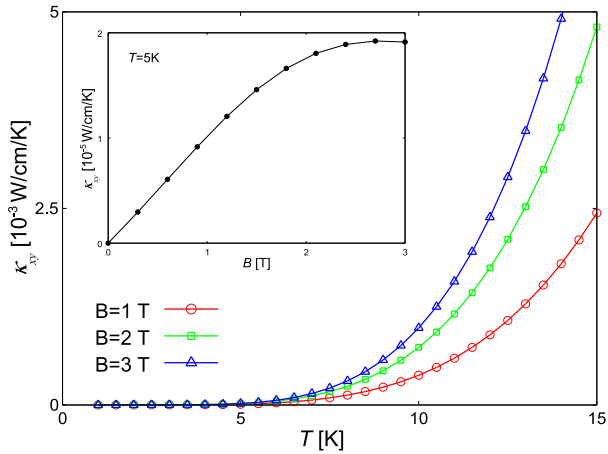}
\caption{Magnetic field dependence of the transverse component of thermal conductivity $\kappa_{xy}$ [10$^{-3}$W/cm/K]. Inset is the magnetic field dependence of $\kappa_{xy}$ [10$^{-5}$W/cm/K] at $T=5$ K. Here, $g=1$ and $\omega_{ab}=0$ and the state $d$ is ignored.      
}
\label{fig3}
\end{figure}

{\it Phonon Hall effect}. --
The Boltzmann equation for the phonon distribution function,
${f_{k}} = f^{(0)}_{k} + g_{k }^{(S)} + g_{k}^{(A)}$, reads~\cite{KL}, 
\begin{equation}
\label{baltzmann}
{c^2}{\bf{k}} \cdot \left( {\frac{{\nabla T}}{T}} \right)
\left( { - \frac{{\partial {f^{(0)}_{\bm k}}}}{{\partial {\omega _{k}}}}} \right) \approx
 \sum\limits_{q } {\left( {W_{\bm q \to \bm k}{f_{\bm q}} - W_{\bm k \to \bm q}{f_{\bm k}}} \right)}.
\end{equation}
Here ${f^{(0)}_{k}}$ is the equilibrium Bose-Einstein distribution.
Since the scattering rate (\ref{w2}) contains both the symmetric part,
$W^{(S)}_{q \to k}= W^{(S)}_{k \to q}$
and the asymmetric part,
$W^{(A)}_{q \to k}=  - W^{(A)}_{k \to q}$, 
we need to account for the two non-equilibrium components, $g_{k}^{(S)}$ and 
$g_{k}^{(A)}$, 
\begin{eqnarray}
g^{(S)}_{k} \propto ({\bm {k}} \cdot {\bm \nabla} T)\ , \ \ \ \ 
g_{k}^{(A)} \propto ({\bm {k}}\cdot[{\bm n}_B\times{\bm \nabla} T]).
\end{eqnarray}
Assuming that the asymmetry parameter in Eq.~(\ref{w2}) is small,
${\cal K}_{\omega}\omega\Gamma_{c\omega} \ll 1$,
solution of the  Boltzmann equation is straightforward and results in the
following non-equilibrium part of the distribution function, 
\begin{eqnarray}
\label{f11}
g_{k }^{(S)} &+& g_{k}^{(A)}
	=-\frac{e^{\omega_k/T}}{(e^{\omega_{k}/T}-1)^2}\frac{c^2}{T^2}
\tau_{\omega}\\
&&\times
\left\{
({\bm k}\cdot{{\bm \nabla}T})
-\frac{1}{3}{\cal K}_{\omega} 
\omega\Gamma_{\omega}
({\bm {k}}\cdot[{\bm n}_B\times{\bm \nabla} T])
\right\}. \nonumber
\end{eqnarray}
Hence, we calculate the diagonal- and the off-diagonal thermal conductivities as,
\begin{eqnarray}
\kappa_{xx}&=&\frac{T^3}{2\pi^2c} \int\tau_{\omega}\frac{x^4e^x dx}{(e^x-1)^2},
\label{ko}\\ 
\kappa_{xy}&=&\frac{T^3}{2\pi^2c} \int \tau_{\omega}\frac{{\cal K}_{\omega}}{3} 
\omega\Gamma_{\omega}\frac{x^4e^x dx}{(e^x-1)^2},\label{ko2}
\end{eqnarray}
where $x\equiv\omega/T$.
The diagonal thermal conductivity in Eq.~(\ref{ko}) is of the standard form~\cite{Berman}, 
which is used to fit the data in Ref.~\onlinecite{inyushkin10}. 
The transverse thermal conductivity $\kappa_{xy}$ given by Eq.~(\ref{ko2}) is shown in Fig.~\ref{fig3} as a function of $T$ with $B$=1,2,3 T. 
%%%%%%%%%%%%%%%%z
%%%%%%%%%%%%%%%%%
We can see that $\kappa_{xy}$ is enhanced by $T$ and $B$ (See also the inset in Fig.~\ref{fig3}).  
Note that this results is justified for $T<\Omega_{a'c}\sim$20 K, since the state $c$ is assumed to be un-populated. The inset in Fig.~\ref{fig3} is the $B$-dependence of $\kappa_{xy}$, which increases and finally starts to decrease around $B\sim$2.5 T.

Our estimate of the phonon Hall angle, $S$, immediately follows from Eqs.~(\ref{ko}) and~(\ref{ko2})
and, for magnetic field larger than 1-2T, is,
\begin{equation}
\label{sc}
S \equiv \frac{1}{B}\frac{\kappa_{xy}}{\kappa_{xx}} %\sim \frac{{\cal K}_{\omega}}{3} \omega \Gamma_{\omega}. 
\end{equation}
Assuming that at temperature $T=5$K the frequency is $\omega=T=5$K, 
Eq.~(\ref{sc}) results in the following estimate: 
$S(T=5K) \sim 5\cdot 10^{-4}{\mbox{rad/T}}$.
An accurate evaluation of the integrals in Eq.(\ref{ko}) confirms that the primary contribution to 
$\kappa_{xx}$ comes from
$\omega \approx T=5$K. On the other hand the dominant contribution to
$\kappa_{xy}$ comes from $\omega \sim 30$K - the phonon Hall effect
is due to relatively ``hot'' phonons. Accounting for the hot phonon effect enhances our theoretical
estimate: 
$S(T=5K) \sim  10^{-3}{\mbox{rad/T}}$.
Our estimate is reasonably consistent with measurements, 
$S(T=5.45K) \approx 1\cdot 10^{-4}{\mbox{rad/T}}$~\cite{strohm}
and
$S(T=5.13K)\approx 0.35 \cdot 10^{-4}{\mbox{rad/T}}$~\cite{inyushkin07}.
The presented theoretical estimates of $\kappa_{xy}$
correspond to $C\sim 1$. 
Important is that $C$-dependence of the Hall angle may explain the significant difference between the two measurements, i.e., two different crystals were used in the two measurements~\cite{strohm,inyushkin07} (see also supplemental material).

{\it Conclusion.}--
We have shown that the puzzling phonon Hall effect observed in Tb$_3$Gd$_5$O$_{12}$, 
is due to  the resonant skew scattering of phonons from the crystal field levels of
superstoichiometric Tb$^{3+}$ ions.
The obtained magnitude of the effect is in agreement with experiments
performed at $T=5$ K.
We predict that the magnitude of the effect  grows very significantly with 
temperature in the interval 3 K $< T <$ 15 K.
Compared to the performed measurements we expect the effect to be about an order of magnitude 
larger at $T=10-15$ K.
A mechanism similar to that considered here for the phonon Hall effect
is also valid for the Hall effect of light~\cite{rikken}: skew scattering
of light from atomic/molecular transitions. For light
the quadrupole crystal field interaction Eq.~(\ref{L1}) should be replaced by 
the electric dipole interaction.

\begin{acknowledgements}
We would like to thank A. I. Milstein, G. Khaliullin and G. Jackeli for stimulating discussions. This work was supported by the Grant-in-Aid for Scientific Research and bilateral program from MEXT. 
M.M. thanks the Godfrey Bequest and the School of Physics at the University of New South Wales for financial support and kind hospitality. 
O.P.S. thanks the Japan Society for Promotion of Science and 
Advanced Science Research Centre JAEA for financial support and kind hospitality. 
\end{acknowledgements}

\section*{SUPPLEMENTAL MATERIALS}
\section{Resonant scattering in thermal conductivity}
The thermal conductivity $\kappa_{xx}$ of Tb$_3$Gd$_5$O$_{12}$ (TGG) 
has been studied by Inyushkin and Taldenkov who measured and analysed the conductivity~\cite{inyushkin10}.
They fit the temperature dependence of $\kappa_{xx}$ by supposing four processes; boundary, point defect, umklapp, and resonant scatterings. It is concluded that 
{\it at helium temperatures for which the phonon Hall effect was detected, $\kappa_{xx}$ is almost completely determined by resonance scattering from impurity ions
and scattering from boundaries (size of the sample)}~\cite{inyushkin10}. 
Following their results, we read the experimental data of Fig.~2 in Ref.~\onlinecite{inyushkin10} and plotted in Fig.~\ref{fig1} by circles (red). 
%%%%%%%%%%%%%%%%%
\begin{figure}[hbt]
\includegraphics[width=0.45\textwidth]{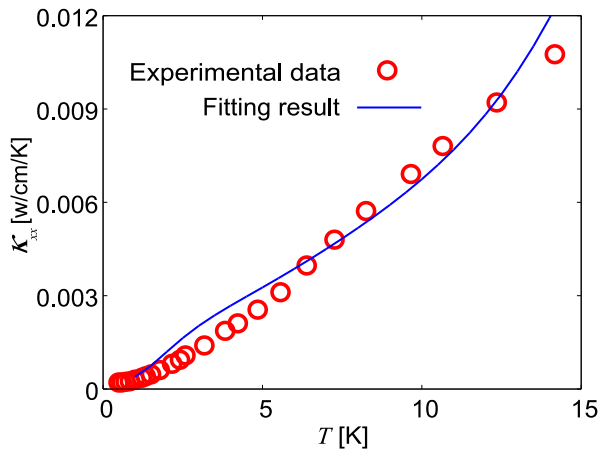}
\caption{The thermal conductivity due to the resonant scattering. 
The circles (red) are obtained by the experimental data (See Fig.~2 in Ref.~\onlinecite{inyushkin10}) and 
the solid line (blue) is our fitting result by the minimal model (See Fig.~1 left in the main text).  
}
\label{fig1}
\end{figure}
%%%%%%%%%%%%%%%%%
To obtain the minimal level scheme at $T<$ 15~K (See Fig.~1 in the main text), we use the following equations, which are equivalent to Eqs.~(4) and (13) in the main text,  
\begin{eqnarray}
&&\kappa_{xx}=\frac{T^3}{2\pi^2c} \int_0^{\omega_D/T}\tau_{\omega}\frac{x^4e^x dx}{(e^x-1)^2},\label{eq1}\\
&&\tau^{-1}_{\omega}= \tau_L^{-1}+\sum_{i=,b,c,d}\tau^{-1}_{ai,\omega},\label{eq2}\\
&&\tau^{-1}_{ai,\omega}=\frac{N_s}{N_{Tb}} \frac{\omega_D^3\omega^4}{80\pi}
 \frac{(\Omega_{ai}/\omega_{ai})^2\Gamma_{ai}^2/\omega_{ai}^4}
{(\omega^2-\Omega_{ai}^2)^2+\Omega_{ai}^2\Gamma_{i\omega}^2},\label{eq3}\\
&&\Gamma_{ai}=\frac{\gamma^2_{i}\omega_{ai}^3}{\pi\rho c^5}=\frac{\gamma^2_{i}\omega_{ai}^3}{\pi\omega_D^3(\rho c^2/40\pi^2 N_{Nb})}\,\nonumber\\
&&\Gamma_{i\omega}=\left(\frac{\omega}{\omega_{aj}}\right)^3\Gamma_{aij} \nonumber,
\end{eqnarray}
where $x\equiv\omega/T$, $c=3.72 \times 10^5$ cm/s, and the mass density $\rho=7.2$ g/cm$^3$~\cite{inyushkin10}. Here $\tau_{L}^{-1}= c/L$ is due to the finite size of the  sample $L\approx 1$mm (boundary scattering).
The total density of Tb ions is $N_{Tb}\approx 1.3\times 10^{22} {\rm cm}^{-3}$, 
the density of superstoichiometric Tb ions is $N_s\approx 1.5\times 10^{20}{\rm cm}^{-3}$, 
and the Debye temperature is 
$\omega_D =487$ K~\cite{inyushkin10}.
Note that in the considered temperature range
the upper limit of integration in Eq.~(\ref{eq1}) is
$\omega_D/T > 30$, so we can safely set it equal to $\infty$.
We found that to fit $\kappa_{xx}$ in the temperature range $T < 15K$ one needs
minimum four levels. Three lowest levels are determined from the fit quite accurately,
$\omega_{ab}=3$ K, $\omega_{ac}=20$ K, $\omega_{bc}=17$ K.
The topmost level, which describes a cumulative effect of all higher states,
is somewhat ambiguous, we take $\omega_{ad}=70$ K, and $\omega_{bd}=67$~K.
The fit with $\gamma_{ab}=1.5$~eV, $\gamma_{ac}=\gamma_{bc}=0.6$~eV, $\gamma_{ad}=\gamma_{bd}=0.8$~eV
is shown in Fig.\ref{fig1} by blue solid line.
The contribution of the topmost $d$-level is relatively small, but still it is important for the fit.
This contribution scales as $\propto \gamma_{ad}^2/\omega_{ad}^2$,
therefore one can always increase $\gamma_{ad}$ and $\omega_{ad}$ proportionally.
Note that since we do not account for thermal population of the $c$-level, our
fit starts to deviate from experimental data at $T > 15K$.

\section{Magnetic field dependence of thermal conductivity}
It is known that the ground state energy doublet is very sensitive to
magnetic field $B$ as shown in the right part of Fig.~1 in the main text, 
$\omega_{ab} \to \Omega_{a'b'}=\sqrt{\omega_{ab}^2+(2gB)^2}$. 
The $\kappa_{xx}$ in a magnetic field calculated at $T=5$~K 
with values of parameters presented above
is shown in Fig.~\ref{fig2}. 
%%%%%%%%%%%%%%%%%
\begin{figure}[ht]
\includegraphics[width=0.4\textwidth]{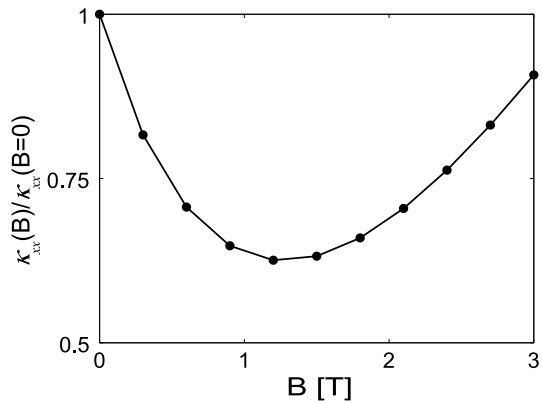}
\caption{Magnetic field dependence of $\kappa_{xx}(B)$ normalized by its magnitude without magnetic field $\kappa_{xx}(B=0)$ . 
}
\label{fig2}
\end{figure}
%%%%%%%%%%%%%%%%%
For simplicity we assume that $|c\rangle$ and $|d\rangle$ are not sensitive to the magnetic field.
The suppression of $\kappa_{xx}$ by a magnetic field is reported in Ref.~\onlinecite{inyushkin10}, and our result is close to the data obtained by the magnetic field in the [111] direction. 
We find that the split of the quasi-doublet state is the main source of such a suppression. 
However, to fit the angle dependence of $\kappa_{xx}$ from Ref.~\onlinecite{inyushkin10}, 
one needs to account different crystallographic positions of superstoichiometric Tb ions with different orientations of crystal fields. In addition, a kind of magnetostriction would be involved in such a magnetic field dependence of $\kappa_{xx}$. 
In the present work we disregard these fine details.

\section{Skew scattering probability}
We already pointed out, Ref.~[27] in the main text, that
$|\pm M\rangle$ states are composed of states with definite z-projection of the ion
angular momentum $J$ such as,  
\begin{eqnarray}
\label{MM}
&&|+M\rangle=...|+2\rangle+\alpha_+|+1\rangle+\alpha_0|0\rangle+\alpha_-|-1\rangle+...\nonumber\\
&&|-M\rangle=...|-2\rangle-\alpha_+|-1\rangle+\alpha_0|0\rangle-\alpha_-|+1\rangle+...\nonumber\\
\end{eqnarray}
The structure of $c$- and $d$-states is similar.
Matrix elements of the interaction Hamiltonian (2) come from transitions with
$\Delta J_z =\pm 1$ and with $\Delta J_z =\pm 2$.
It is easy to check that in the scattering amplitude Fig.~2 the contributions with
$\Delta J_z =\pm 1$ result in $\cos\phi$ or $\sin\phi$ and
the contributions with $\Delta J_z =\pm 2$ result in $\cos2\phi$ or $\sin2\phi$.
Therefore, the scattering probability reads
\begin{eqnarray}
W_{\bm k\to \bm q}^{a'c}
&\propto&
\left\{ A_1\left[\cos\phi  - \frac{\omega \Gamma_{c \omega}}{2\Omega_{a'c}^2}\sin \phi\right]
\right.
\nonumber\\
&+&\left.
A_2\left[\cos2\phi  - \frac{\omega \Gamma_{c \omega}}{2\Omega_{a'c}^2}\sin 2\phi\right]
\right\}^2\ ,\label{w1}
\end{eqnarray}
In Eqs.~(5) and (6) in the main text for simplicity we set $A_2=0$.
Generally, while one expects some suppression of the second
harmonic, $A_2/A_1 < 1$, the harmonic is of course nonzero.
A particular value of $A_2/A_1$ depends on the coefficients
in the wave functions (\ref{MM}).
Both the $A_1^2$-term and the $A_2^2$-term in (\ref{w1}) have the $\phi \to \pi-\phi$ compensation
problem. These terms, as it is described in the main text, contribute to PHE only
due to correlation of positions of impurities. This contribution is proportional to the 
correlation coefficient $C$ in Eq.(9). 
We think that this is the leading term dominating the skew scattering and can explain the sample dependence of PHE.
However, there is also the $A_1A_2$ interference term
in (\ref{w1}). The interference term results in the forward-backward scattering asymmetry
and hence the terms does not have the $\phi \to \pi-\phi$ compensation problem.
This interference contribution in Eq.(9) is {\it not }proportional to the impurity
correlation coefficient $C$. 
Hence, the skew scattering is always there, even if the impurity correlation $C$ was ignored. The term proportional to $A_2/A_1>0$ implies that the forward phonon
scattering dominates over the backward one, which is intuitively natural. 
In this study, we consider that the contribution of $A_2/A_1$-term will be smaller than that of $C$-term to the PHE. Detail analysis on the electronic states of TGG will judge this point in the future.

\section{Derivation of Eq. (9) in the main text}
In the main text we point out that there are two mechanisms which destroy
$\phi\to \pi-\phi$ compensation: (i) Spatial correlation of impurity 
positions; (ii) Interference between contributions with 
different values of $\Delta J_z$. Here we show how the mechanism (i) works.
Eq. (\ref{w1}) results in the following skew terms
\begin{eqnarray}
W_{\bm k\to \bm q}^{(1)}
&\propto& ({\bm n}_k\cdot{\bm n_q})({\bm n}_B\cdot[{\bm n}_k\times {\bm n}_{q}])
\label{w2}\\
W_{\bm k\to \bm q}^{(2)}
&\propto& \frac{A_2}{A_1}({\bm n}_B\cdot[{\bm n}_k\times {\bm n}_{q}]) \ .
\label{w3}
\end{eqnarray}
Here we neglect small $A_2^2$ terms and $A_2/A_1$ term is discussed above. 
The multiplier $({\bm n}_k\cdot{\bm n_q})$ in $W^{(1)}$ leads to the
$\phi \to \pi-\phi$ compensation in $\kappa_{xy}$.
It is discussed in the main text that the spatial correlation of impurities
transforms $W^{(1)}$ due to $P_{\phi}$ such as,
\begin{eqnarray}
W_{\bm k\to \bm q}^{(1)}&\to& P_{\phi}W_{\bm k\to \bm q}^{(1)} \to
Ca_1(\omega) ({\bm n}_k\cdot{\bm n_q})^2
({\bm n}_B\cdot[{\bm n}_k\times {\bm n}_{q}]) \ .\nonumber\\
\label{w4}
\end{eqnarray}
This expression does not contain the $\phi \to \pi-\phi$ compensation.
Now we want to reduce (\ref{w4}) to the standard skew correlation
$({\bm n}_B\cdot[{\bm n}_k\times {\bm n}_{q}])$ used in Eq.(9) in the main text.
A naive way is just to replace $({\bm n}_k\cdot{\bm n_q})^2 \to 1/2$.
However, this is an obvious overestimation, the correlation
$({\bm n}_B\cdot[{\bm n}_k\times {\bm n}_{q}]) =\sin\phi$ is maximum
at $\phi=\pi/2$ where $({\bm n}_k\cdot{\bm n_q})^2=(\cos\phi)^2$ is zero.
The correct way is to substitute (\ref{w4}) in Eq.(10),
solve the kinetic equation with respect to $g^{(A)}_k$ defined in Eq.(11),
and finally map the solution back to the simple skew correlation 
$({\bm n}_B\cdot[{\bm n}_k\times {\bm n}_{q}])$. 
It is noted that most of term in Eq.(11) substituted by (\ref{w4}) and $g_k^{(A)}$ disappears due to the symmetry, e.g., $\int d\Omega_q g_q^{(A)}=0$ and so on. 
The solution of kinetic equation contains averaging of the fourth rank tensor
$\langle n_{\alpha}n_{\beta}n_{\gamma}n_{\delta}\rangle$, which gives the 
factor 1/5 (see also Eq.~(\ref{average})). All in all, this procedure leads to the term $\frac{C}{5}a_1(\omega)$ in the square brackets in the
second line of Eq.~(9).

\section{Derivation of Eq. (12) in the main text}
Quantities which enter in the r.h.s of Eq.~(10) in the main text
are of the following form
\begin{eqnarray}
W_{\bm k\to \bm q} &=& \frac{\tau_{\omega}^{-1}}{4\pi} \left\{1  - 
a\cdot
{\bm n_B}\cdot[{\bm n_k}\times{\bm n_q}]\right\}, \label{eq7}\\
f_{\bm k}&=&f_{\bm k}^{(0)}+A^S{\bm n_{\bm k}}\cdot\nabla T+A^A{\bm n_{\bm k}}\cdot[{\bm n_{\bm B}}\times\nabla T],\nonumber
\end{eqnarray}
where the coefficient $a$ defined in Eq. (9) is small, $a\ll 1$. 
Using Eqs.~(\ref{eq7}) we find the r.h.s of Eq.~(10) in the main text,
\begin{eqnarray}
&& \sum\limits_{q } {\left( {W_{\bm q \to \bm k}{f_{\bm q}} - W_{\bm k \to \bm q}{f_{\bm k}}} \right)} \label{eq23}\\
&&=-\frac{\tau_{\omega}^{-1}}{4\pi} \sum_{\bm q} \left\{
A^S{\bm n_{\bm k}}\cdot\nabla T
+A^A{\bm n_{\bm k}}\cdot[{\bm n_{\bm B}}\times\nabla T] \right.\nonumber\\
&&\left. +a\cdot A^S({\bm n_B}\cdot[{\bm n_q}\times{\bm n_k}])({\bm n_{\bm q}}\cdot\nabla T)
\right\}\nonumber\\
&=&-\tau_{\omega}^{-1} \left\{
A^S{\bm n_{\bm k}}\cdot\nabla T
+(A^A+\frac{a}{3}A^S){\bm n_{\bm k}}\cdot[{\bm n_{\bm B}}\times\nabla T] 
\right\},\nonumber
\end{eqnarray}
When calculating  (\ref{eq23}) we neglect terms $\propto a^2$
and keep in mind that
\begin{eqnarray}
&&\int d\Omega_{\bm q} 1=4\pi\nonumber\\
&&\int d\Omega_{\bm q} n_{{\bm q}\mu} n_{{\bm q}\nu}=\frac{4\pi}{3}\delta_{\mu\nu}\ , \label{average}
\end{eqnarray}
where $\mu$ and $\nu$ are Cartesian indexes and $\Omega_{q}$ is solid angle.
Comparing (\ref{eq23}) with l.h.s of Eq.~(10) we find the textbook expression
for $A^S$ and, we also find the condition $A^A=-(a/3)A^S$.
Hence we come to Eq.~(12)
\begin{eqnarray}
&& g_k^{(S)}+g_k^{(A)}=A^S\left({{\bm n_{\bm k}}\cdot\nabla T}-\frac{a}{3}{\bm n_{\bm k}}\cdot[{\bm n_{\bm B}}\times\nabla T]\right)\nonumber\\
&& A^S=-\frac{e^{\omega_k/T}}{(e^{\omega_{k}/T}-1)^2}\frac{c^2}{T^2}
\tau_{\omega}k \ .\nonumber
\end{eqnarray}


\begin{thebibliography}{00}
\bibitem{note3}
Throughout this paper, we set both Planck's constant and Boltzmann's constant equal to unity: $\hbar=k_B=1$.
\bibitem{note4}
The $|\pm M\rangle$ states are composed of states with definite z-projection of the ion
angular momentum $J$ such as,  $|+M\rangle=...|+2\rangle+\alpha_+|+1\rangle+\alpha_0|0\rangle+\alpha_-|-1\rangle+...$, and $|-M\rangle=...|-2\rangle-\alpha_+|-1\rangle+\alpha_0|0\rangle-\alpha_-|+1\rangle+...$
\bibitem{note1} 
In the skew part we account only for the virtual c-state and
neglect the virtual d-state. This is because
$\Gamma_{d\omega}/\Omega^2_{ad}\ll \Gamma_{c\omega}/\Omega^2_{ac}$
and hence the skew component for the d-state is relatively small.
\end{thebibliography}
\end{document}